\documentstyle[prl,aps,epsfig,floats,twocolumn]{revtex}
\begin{document}
\draft
\twocolumn[\hsize\textwidth\columnwidth\hsize\csname
@twocolumnfalse\endcsname

\title{Efficiency and persistence in models of adaptation}
\author{R. D'Hulst and G.J. Rodgers}
\address{Rene.DHulst@brunel.ac.uk and G.J.Rodgers@brunel.ac.uk}
\address{Department of Mathematical Sciences, Brunel University}
\address{Uxbridge, Middlesex, UB8 3PH, UK}
\date{\today}

\maketitle
\widetext
\begin{abstract}
A cut-and-paste model which mimics a trial-and-error process of 
adaptation is introduced and solved. The model, which can be thought of as a diffusion process with memory, is characterized by two properties, efficiency and persistence. We establish a link between these properties and determine two 
transitions for each property, a percolation transition and a depinning transition. If the adaptation process is iterated, the antipersistent state becomes an attractor of the dynamics. Extensions to higher dimensions are briefly discussed.

\vspace{0.2cm}
\vspace{0.2cm}
\end{abstract}

]
\narrowtext

For some time, physicists have been working on problems outside the scope of traditional physics, as illustrated by recent papers on the cooperation between actors \cite{wasserman}, the repetition of ancestors in genealogical trees \cite{derrida} or the scaling of football goal distributions \cite{malacarne}. 
In many of these systems, whether they are in economics, the social sciences or biology, the process of adaptation, or how an agent or organism adapts to a 
particular event, plays a very important role. Unfortunately there is no general theory of adaptation which reflects either the ubiquity or the simplicity of the process. In this Letter, we introduce and solve toy models for adaptation, in order to identify some basic properties of the process. We investigate the dynamics of a trial-and-error adaptation process, which can be represented as diffusion with memory. First, a random guess is made, then, the system reacts to this initial guess. After this two-step process, it is interesting to consider if the adaptation process has improved the original random guess or not. To answer this question, a measure of efficiency of the adaptation is necessary, and we will show that this property, efficiency, is closely related to another property of the adaptation process, persistence.

Although we will not explicitly model microscopic elements, it is 
helpful to suppose that there exists some microscopic adaptive entities, generically called agents. These agents are trying to guess what the best action would be and the result of their aggregate decisions is a macroscopic parameter, $x$. After this initial step, the agents detect a global result which is the difference between $x$ and 1/2. The individual reactions of the different agents to this information produces a diffusion of size $p$ from the larger population to the smaller one. The ideal reaction would result in $x=1/2$. However, they lack a leader who can synchronize their individual choices and achieve perfect coordination. If there is cooperation, it has to emerge spontaneously from the system. In this paper, we do not concern ourselves with the precise details of the implementation of the agents' decision process, nor on how they adapt. The only questions we investigate are the influence on efficiency of the size of the adaptive population, and how efficiency is related to persistence.

\begin{figure}[h]
\centerline{\psfig{file=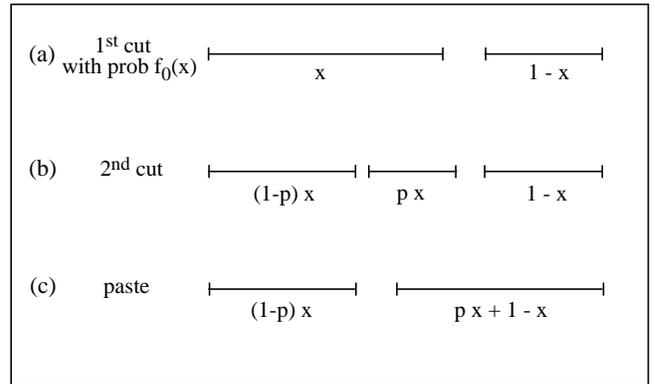,width=8.5cm}}
\caption{Schematic view of the cut-and-paste model, where we take $x> 1/2$, without loss of generality.}
\label{fig1:themodel}
\end{figure}

Consequently, the model is defined as follows. In the first step, a quantity of length 1 is cut into two pieces of length $x$ and $1-x$ with probability $f_0(x)$, as shown in Fig. \ref{fig1:themodel}~(a).  In the second step, a fraction $p$ of the larger piece is cut and pasted to the smaller piece, as shown in Fig. \ref{fig1:themodel}~(b) and (c), respectively. This transforms the random guess distribution $f_0 (x)$ into an `educated guess' distribution $f_1 (x)$. In the model, we imagine that we are macroscopic witnesses of some microscopic adaptation process, which is not modeled explicitly. The object of the model is to investigate the appearence of cooperation between microscopic entities in the absence of global control. Here, cooperation is synonymous with obtaining two pieces of equal length. After the two step process, a record of the improvement in cooperation is made by updating a quantity $A$. $A \to A + 
1$ when $x$ is closer to 1/2 after than before adaptation and $A \to A - 1$ otherwise, with $A\ge 0$. We assume that the model is played several times, without any correlations between the different realizations, apart from the fact that we always start with the same distribution $f_0 (x)$. $A$ is a biased measure of the rate of increase in cooperation after adapting. Adaptation is driven by the prospect of improving on an initial situation, which keeps $A$ non-negative.

For simplicity, we consider that $f_0 (x) = f_0 (1-x)$, implying that 
$f_1 (x) = f_1 (1-x)$, where $f_1 (x)$ is the size distribution after 
adaptation. This distribution is given by

\begin{eqnarray}
\nonumber
f_1 (x) &=& \int_0^{1/2} dy\ f_0 (y)\delta (y+p(1-y)-x)\\
  && + \int_{1/2}^1 dy\ f_0 (y) \delta (y(1-p) -x)
\label{eq:second-cut}
\end{eqnarray}
with a mean value $m_1 = 1/2$. The magnitude of the deviations from 
$m_1$ determine the efficiency of the system. There are several 
quantities capable of measuring these deviations and they all give 
qualitatively similar results. We choose to use the variance 
$\sigma^2_1$ of $f_1 (x)$, equal to

\begin{equation}
\sigma^2_1 = (1-p)^2 \sigma^2_0 + \frac{p(3p-2)}{4} + 2 p (1-p) K_0
\label{eq:variance}
\end{equation}
where $\sigma_0^2$ is the variance of $f_0(x)$, and we define

\begin{equation}
K_0 \equiv \int_0^{1/2} x f_0 (x)\ dx.
\end{equation}
For this measure of the efficiency, the optimal value of the adaptation is given by the minimum of $\sigma_1^2$,

\begin{equation}
p_c = \frac{4 \sigma^2_0 -  4 K_0 + 1}{4 \sigma^2_0 - 8 K_0 + 3}.
\label{eq: critical p}
\end{equation}
The boundary  between doing better and doing worse when adapting is 
determined by $\sigma_1^2 (p^{*})= \sigma_0^2$, and this gives $p^{*} = 
2 p_c$, independent of $f_0 (x)$.

However, as we mentioned earlier, we assume that there are no correlations 
between successive realizations of the model. Hence, $\sigma_1^2$ represents a 
statistical measure of efficiency over several realizations of the 
model, but this is a quantity that is not available to anyone seeing 
the game played once. In other words, if we imagine an agent taking 
part in the cut-and-paste model, his measure of efficiency is based 
on one realization of the model. The 
probability $\Gamma (p)$ of improving on an initial guess is equal to

\begin{equation}
\Gamma (p) = 2 \int_0^{(1-p)/(2-p)} f_0 (x)\ dx,
\end{equation}
which is a monotonically decreasing function of $p$, equal to 1 when 
$p=0$ and 0 when $p=1$. Like $\sigma_1^2$, $\Gamma (p)$ is not a 
property directly available after one realisation of the model, but 
it gives a probability about the outcome of one such realization. If 
we remember that, after adapting, we want to make a judgement upon whether adaptation was worth it or not, $\Gamma (p_{oa}) = 0$ corresponds to 
the onset of adaptation. That is, for $p< p_{oa} = 1$, it becomes 
possible that adaptation leads to an improvement, which is similar to a 
percolation transition \cite{stauffer}. The order parameter of the 
transition is $\Gamma (p)$ which scales as $\Gamma \sim 
(p_{oa}-p)^{\beta}$ for $p$ less than, but close to, $p_{oa}$. The 
value of the critical exponent $\beta$ depends on the analytical form 
of $f_0 (x)$. For $\Gamma (p_{an}) = 1/2$, the system undergoes 
another transition, from non-adaptive to adaptive. For $p> p_{an}$, 
$A$ stays close to 0, while for $p< p_{an}$, $A$ is `depinned' from 0 
and goes away at a velocity $v_{an} = 1 - 2 \Gamma (p)$. This 
velocity, equal to 0 for $p>p_{an}$, is the order parameter of the 
transition \cite{barabasi}. For $p<p_{an}$, the symmetry between 
adapting or not is broken.

\begin{figure}[h]
\centerline{\psfig{file=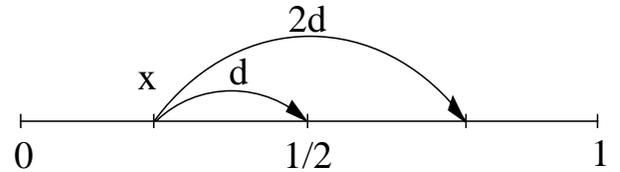,width=8.5cm}}
\vspace{-4.5cm}\caption{Relation between the value for the best 
adaptation ($d$) and the value to obtain the symmetric situation with 
respect to $1/2$ ($2d$). This relation is at the origin of the 
connection between persistence and efficiency.}
\label{fig2:symmetry}
\end{figure}

The two previous quantities, $\sigma_1^2$ and $\Gamma (p)$, are both 
related to the efficiency. Another important property of the model is its 
persistence, which measures the propensity of one piece being the smaller piece both before and after adaptation. The probability that the smaller piece 
remains the smaller piece after adaptation, the persistence 
probability $\Pi (p)$, is equal to
\begin{equation}
\Pi (p) = 2 \int_{0}^{(1-2p)/2(1-p)} f_0 (x)\ dx
\end{equation}
if $p<1/2$ and zero otherwise. In fact, persistence and efficiency are 
closely related quantities, as can be seen in Fig. 
\ref{fig2:symmetry}. For any position $x$ at a distance $d$ from 1/2, 
the symmetric position with respect to 1/2 is at distance $2d$ from 
$x$. In other words, if $d$ marks the boundary between 
antipersistent and persistent, then $2d$ is at the boundary 
between better and worst than the previous step. This relation is expressed 
mathematically by the fact that similar properties for efficiency and 
persistence are characterized by $\Gamma (p_{eff}) = \Pi (p_{pers})$, 
which implies $p_{eff} = 2 p_{pers}$. For instance, the onset of persistence and adaptation are characterized by $\Pi (p_{op}) = \Gamma (p_{oa}) = 0$. From the previous relation, one obtains the relations $p_{oa} = 2 p_{op}$, 
where $p_{op}$ signals the onset of persistence, $\Pi (p_{op}) = 0$ 
and $p_{an} = 2 p_{ap}$, where $p_{ap}$ signals the antipersistent-persistent transition, $\Pi (p_{ap}) = 1/2$. Hence, a characteristic property of persistence in the system requires twice as much reaction from the system to have the same effect on the system efficiency. Note that the antipersistent-persistent transition corresponds to a symmetry breaking between left and right for $p< p_{ap}$.

\begin{figure}[h]
\centerline{\psfig{file=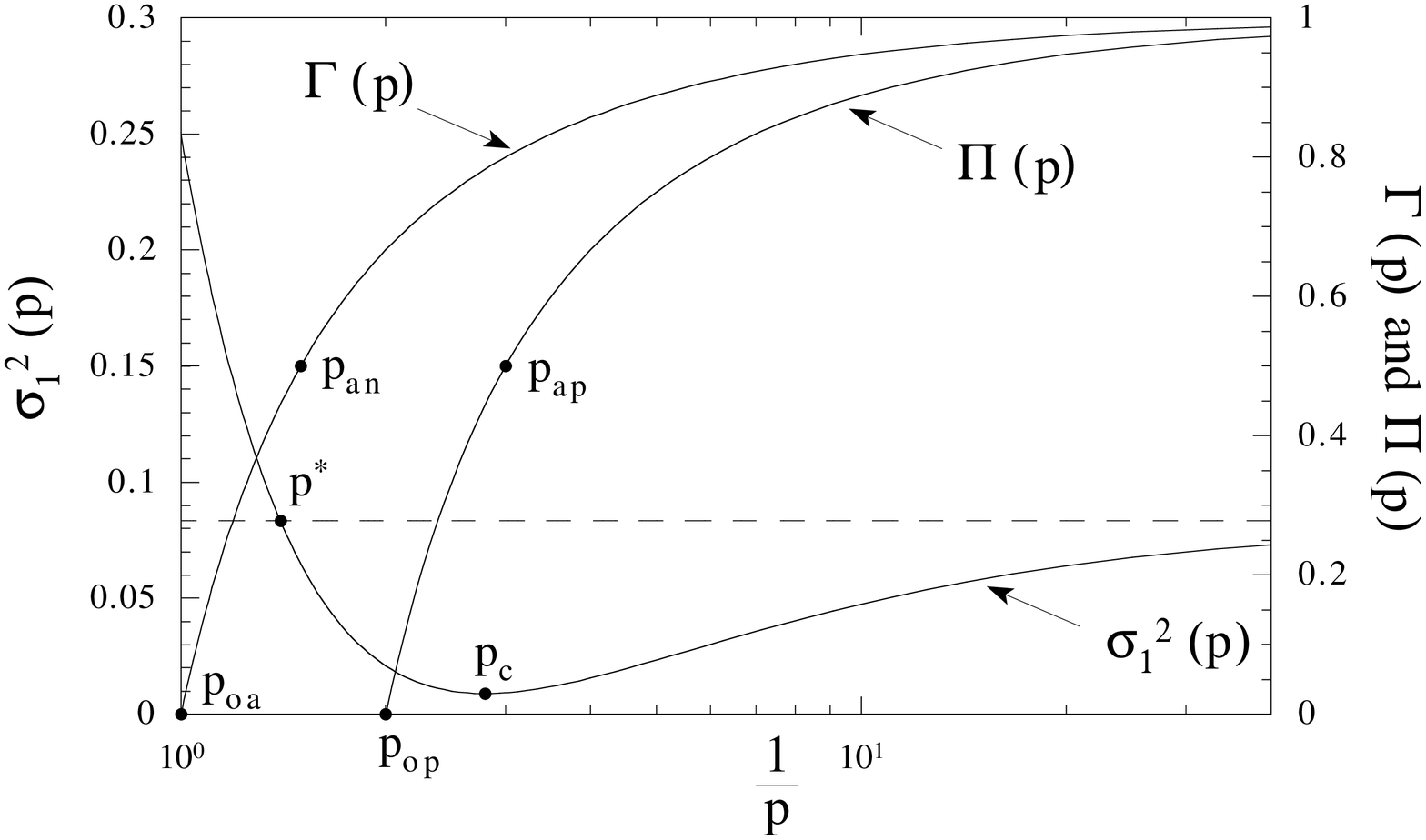,width=8.5cm}}
\caption{Variance $\sigma_1^2$ of $f_1$ as a function of $p$ (left 
scale), persistence probability $\Pi$ and improvement probability 
$\Gamma$ as functions of $p$ (both on the right scale). The initial 
distribution $f_0 (x)$ is uniform on $(0,1)$. The dots ($\bullet$) 
mark the special values of $p$ considered in the text. The dashed 
horizontal line shows the variance of $f_0 (x)$.}
\label{fig2:uniform}
\end{figure}

We illustrate the previous results in Fig. \ref{fig2:uniform} with a 
uniform distribution $f_0 (x)$. With this choice, $p_{ap} = 1/3$, 
$p_c= 5/14$, $p_{op} = 1/2$, $p_{an} = 2/3$, $p^{*} = 5/7$ and 
$p_{oa} = 1$. We have chosen to show the different functions as 
functions of $1/p$ to allow a visual comparison with the Minority 
Game, as we explain later.

Assuming very strict bounded rationality \cite{conlisk}, we have considered up to now that the system was adapting once only. However it is straightforward to apply the cut-and-paste model recursively. After an initial cut with distribution $f_0 (x)$, a fraction $p$ of the larger piece is cut and pasted on the side of the smaller piece, creating a new distribution $f_1 (x)$. Then, a fraction $p$ of the larger piece is cut and pasted on to the side of the smaller piece, creating $f_2 (x)$, and so on. The main conclusion is that the system can be persistent for a transient period, depending on the initial conditions, but that it will eventually enter an antipersistent phase where the left piece is alternatively the larger and then the smaller piece. For a given value of $p$, the system has entered the antipersistent attractor for all initial conditions after

\begin{equation}
i_{att} = \frac{\ln \frac{1}{2}}{\ln (1-p)} - 1
\end{equation}
adaptations. We find that

\begin{equation}
\lim_{i\to \infty} f_i (x) = \frac{\delta \left( x - \frac{1}{2-p} 
\right) + \delta \left( x - \frac{1-p}{2-p}\right) }{2},
\end{equation}
and the system becomes fully efficient only for quasistatic 
adaptation, with the larger and the smaller pieces ultimately becoming 
equal if $p \to 0$.

\begin{figure}[h]
\centerline{\psfig{file=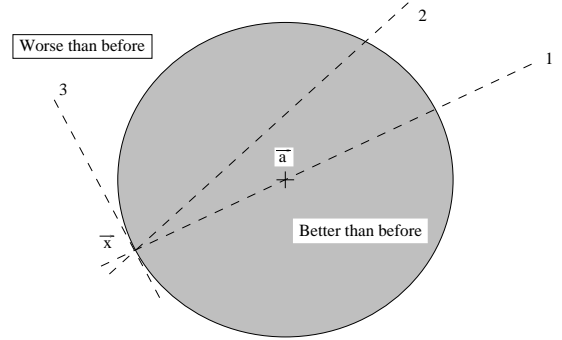,width=8.5cm}}
\caption{Adaptation on a two-dimensional plane. If the adaptation is 
along 1 or 2, the model is similar to a one-dimensional model, while 
for adaptation along 3, there is no better-than-random phase.}
\label{fig3:2d}
\end{figure}

Finally let us mention that the cut-and-paste model can easily be 
extended to higher dimensions, using vectors $\vec{x}$ and $\vec{p}$ 
as the random guess and adaptation respectively, and another point 
$\vec{a}$ as the implicit aim. We present a two-dimensional example in 
Fig. \ref{fig3:2d}. As seen in the figure, depending on $\vec{p}$, 
the adaptation is either better or worse than random, the best 
possible adaptation being along line $1$, a better than random 
adaptation can also be achieved along line 2, while along line 3, the 
adaptation process is always inefficient. The circle marks all points 
that are equivalent to $\vec{x}$. If  $\vec{x}$ is the initial guess, 
falling inside the circle after adaptation corresponds to improving 
on the initial guess.

In summary, we have presented a cut-and-paste model to mimic a 
trial-and-error process of adaptation, and showed that the model 
displays two properties, persistence and efficiency. A pair of 
transitions is associated to each property. A pair is composed of a 
percolation transition, corresponding to the onset of the property, 
and a depinning transition, corresponding to the growth of the 
corresponding property. One particular point to underline is the 
close relationship between efficiency and persistence. A 
characteristic property of persistence in the system requires twice 
as much reaction from the system to have the same effect on the 
system efficiency. If the model is iterated, the system always 
reaches an antipersistent state, where the left piece is alternately 
smaller and then larger. 
Our model is inspired by the Minority Game, 
a model to mimic the exchange of a commodity between agents which, 
despite its more complex features, displays transitions similar to 
the ones presented here \cite{cz,smr}. We will provide a detailed 
comparison with the Minority Game elsewhere \cite{dr}, but here we presented 
Fig. \ref{fig2:uniform} as a function of $1/p$ instead
 of $p$ to 
allow a visual comparison with the variance of the number of buyers in the Minority Game \cite{smr}. 
Numerous properties found here are
reminiscent of the Minority Game. For instance, while much of the 
analytical research has concentrated on the system efficiency, it is the 
antipersistent-persistent transition which is best understood, thanks 
to an analogy with spin glasses \cite{cmz}. 
Another point to note is 
that the critical parameter of the Minority Game, a parameter called 
$\alpha$, has two important values,  $\alpha_c$ where the variance is a 
minimum and  
$\alpha^{*}$ where the variance is equal to a random guess. 
These are related by $\alpha^{*}\approx\alpha_{c}/2$. 
In fact, we argue that the original Minority Game is 
itself an intricate formulation of this simple mechanism, where the size of the adapting population is a complex quantity depending on the history given to 
the agents and the strategy space. The simplicity of 
the cut-and-paste model suggests that many of the more complex features 
of the Minority Game may be more common and universal than originally 
believed. 

One of us (GJR) would like to thank The Leverhulme Trust for financial support.

\end{document}